\documentstyle[12pt,epsf]{article}
\newlength{\myleftmargin}
\newlength{\paperwidth}
\begin{document}
\thispagestyle{empty}
\baselineskip 18pt
\begin{flushright}
  {\tt DPNU-95-12 \\ June 1995}
\end{flushright}
 \vspace{0.1em}
\begin{center}
   {\Large {\bf Finite $N_c$ Results for $F/D$ Ratios of}}
\vskip 1mm 
   {\Large {\bf the Baryon Vertices and $I=J$ Rule}}
\end{center}
  \vspace{0.3em}
\begin{center}
  {\sc Akira~~Takamura},~~~{\sc Shoji~~Sawada},~~~{\sc Yoshimitsu~~Matsui}\\
  and~~~{\sc Shinsaku~~Kitakado}\footnote[1]{
e-mail addresses: $<$takamura@eken.phys.nagoya-u.ac.jp$>$,
 $<$sawadas@eken.phys.nagoya-u.ac.jp$>$,
 $<$matsui@eken.phys.nagoya-u.ac.jp$>$ and 
 $<$kitakado@eken.phys.nagoya-u.ac.jp$>$}

\end{center}
\begin{center}
  \sl{Department of Physics, Nagoya University,}\\
  \it{Nagoya 464-01, Japan}
\end{center}
\vskip 1.5em  
{\large {\bf Abstract~:}}
 We calculate the $F/D$ ratios of spin-nonflip
 baryon vertex for an arbitrary number of color degrees of freedom $N_c$
 both in the non-relativistic quark model
  with the  $SU(6)$
 spin-flavor symmetry and in  the chiral
 soliton model with  $SU(3)$ flavor symmetry. 
 We find that the spin-nonflip $F/D$ ratio tends to $-1$ in the limit of
 $N_c \to \infty$.
 We show that this leading value  $F/D= -1$ of spin-nonflip baryon vertex
  in the $1/N_c$ expansion corresponds to the isoscalar dominance
  while the well known leading value $F/D=1/3$ of the 
 spin-flip vertex corresponds to the isovector  dominance. We discuss
  origins  of the dominance of isovector in spin-flip and isoscalar in
spin-nonflip baryon vertices, referred to as the $I=J$ rule. 
  \par
In terms of the matrix elements of the operator
 which transform as the generator $\lambda^8$  
of the $SU(3)$ symmetry we  derive the model independent
 isoscalar formula for baryon vertices and  apply this
 to the mass formula and the isoscalar part of the baryon magnetic moments.
 The same Okubo-Gell-Mann mass relation and its refined relation
among the octet baryons   
 as the one for the case $N_c=3$  is derived model independently for
 arbitrary  color degrees of freedom $N_c$.
 Contrary to $\lambda^8$,
 the Coleman-Glashow mass relation which is derived from an operator that
 transform as  $\lambda^3$ of $SU(3)$
 symmetry only hold for $N_c =3$.
 \vskip 1.5cm 
{\quad \quad}



\section{Introduction}
\par 
The $1/N_c$ expansion was proposed as a non-perturbative approach to QCD 
 by 't~Hooft in 1974$\cite{tHooft}$. He has shown that the Feynman diagrams
 which are relevant at the leading order of the $1/N_c$ expansion are the
 planar diagrams without internal quark loops.
  \par
 Based on this $1/N_c$ expansion 
  Witten suggested that in the large $N_c$ limit a
  baryon looks like soliton$\cite{Witten}$. 
  {}From this viewpoint the Skyrme's conjecture that baryons are
  solitons of the 
  nonlinear chiral Lagrangian for the chiral fields has been revived in
  early 1980's and has succeeded in describing the baryon sector 
  from the meson  sector at least semi-quantitatively$\cite{Nappi}$.
  \par
 The non-relativistic quark model(NRQM), on the other hand,
   has successfully described
  the hadron phenomena in various aspects and played a crucial role
  in the way of establishing QCD. However the NRQM description of the low 
  energy hadron physics has not been derived from QCD
  and looks to have different origin compared with   
    the chiral soliton model(CSM).
  \par
  The $1/N_c$ expansion method has been considered to be a qualitative
  method to study QCD.
 Recent  extensive studies of the consistency conditions
$\cite{Gervais} \cite{Dashen} \cite{Dashen2}$
 show that the $1/N_c$ expansion method of QCD is useful for the analysis of
 NRQM and CSM from model independent viewpoint.
 \par
 In the previous paper$\cite{Takamura}$ we have calculated the $F/D$ ratios
 of spin-flip baryon vertices, denoted hereafter as $F_-/D_-$,
both  in the NRQM and the CSM
  for arbitrary color degrees of freedom $N_c$.
  The value of $F_-/D_-$ 
   tends to $1/3$ in the large $N_c$ limit and the physical meaning of this
 value was nothing but the isovector dominance for the spin-flip baryon
  vertex.
  \par
  In this paper we study the spin-nonflip baryon vertices
   for arbitrary  $N_c$  both in the NRQM and the CSM and calculate
   the $F/D$ ratio of the spin-nonflip baryon vertices, denoted
   as $F_+/D_+$. The $F/D$ ratios are considered to reflect 
   characteristics of QCD that do not depend on details of the special
    effective
    models of QCD.
   We discuss the obtained results  and compare with those of
    the spin-flip baryon  vertices from the viewpoint of $1/N_c$ expansion
    and consistency conditions.
    \par
 In \S 2, we define the baryon state for arbitrary 
 $N_c$. In \S 3 and \S 4,  we calculate 
  the $F/D$ ratios of spin-nonflip baryon vertex
   for arbitrary $N_c$ both in  the NRQM and
  the CSM.
  In \S 5, we discuss  the implications of
 $F_-/D_-=1/3$ and $F_+/D_+=-1$ and the convergence problem of $1/N_c$ 
 expansion.
  In \S 6, we derive the isoscalar
 formula for arbitrary $N_c$. We  also comment on   the
 mass formula which was derived by Dashen, Jenkins and Manohar$\cite{Dashen}$.
We show there that the
 same Okubo-Gell-Mann mass relation and its refined relation
among the octet baryons   
 as the one for the case $N_c=3$  is derived model independently for
arbitrary $N_c$ while
 the Coleman-Glashow mass relation is derived only for $N_c =3$.
 \par

\section{The Large $N_c$ Baryon States}
\hspace*{\parindent}
In order to study the  properties of baryons in the $SU(N_c)$
symmetric QCD with arbitrary $N_c$
we have to introduce the extended baryon state which is totally antisymmetric
 color singlet state of the $SU(N_c)$ symmetry.
In the case of two flavors with the $SU(2)$ symmetry
the extensions from both NRQM and CSM are simple, 
because the spin wave functions
are the same as those of flavor giving the $I=J$ structure
\begin{equation}
 I=J=\frac{1}{2},~\frac{3}{2},~\frac{5}{2},\cdots,\frac{N_c}{2}.
\end{equation}
 This is a
consequence of the fact that spin-flavor states of the ground state baryon
are totally symmetric in NRQM and is similar to that
of the hedgehog Ansatz in CSM, i.e. $I=J=1/2,~3/2,\cdots,~N_c/2,\cdots$.
In both models with flavor $SU(2)$,
 the baryon ground state with spin 1/2 belong to the isospin
 1/2 state.

In the case of three or more flavors there are some ambiguities
 in  extending  the baryons for large $N_c$ and we have to introduce 
 some unphysical members of the $SU(N_f)$ multiplet.
  However, we will show that
the following special choice of extension for the large $N_c$
 baryon is appropriate to obtain the correct 
  large $N_c$ behavior of various baryon
 matrix elements.
 \par
   The ground state spin 1/2 baryons for arbitrary  $N_c$ belong
 to $(1+k)(3+k)$ dimensional representation of
 flavor $SU(3)$ symmetry because of total symmetry in the spin-flavor
 states where $k = (N_c -1)/2$ $(k=0,1,2,\cdots)$.
  This representation is specified by the
 Young diagram with the first row of length $k+1$ and the second row of
 length $k$ and the root diagram shown in Fig.~1.
The usual octet baryons are located at the top region of this root diagram.
 The flavor wave functions are represented by
 the tensors with one superscript and $k$ subscripts and 
 the usual octet baryons are given by$\cite{Dashen}$
\begin{eqnarray}
& & p~:~~ B^1_{3,\cdots,3}=1 ~~~~ n~:~~ B^2_{3,\cdots,3}=1 \label{N}\\
& & \Sigma^+~:~~ B^1_{23,\cdots,3}=\frac{1}{\sqrt{k}} ~~~~~
    \Sigma^-~:~~ B^2_{13,\cdots,3}=-\frac{1}{\sqrt{k}} \label{Sigma1}\\
& & \Sigma^0~:~~ B^1_{13,\cdots,3}=-B^2_{23,\cdots,3}=
    -\frac{1}{\sqrt{2k}} \label{Sigma2}\\
& & \Lambda~:~~ B^1_{13,\cdots,3}=B^2_{23,\cdots,3}=
     \frac{1}{\sqrt{4+2k}} ~~~~
    B^3_{3,\cdots,3}=-\frac{2}{\sqrt{4+2k}} \label{Lambda}\\
& & \Xi^0~:~~ B^3_{23,\cdots,3}=-\frac{3}{2}B^2_{223,\cdots,3}=
    -3B^1_{123,\cdots,3}=\sqrt{\frac{3}{k(k+2)}} \label{Xi1}\\
& & \Xi^-~:~~ B^3_{13,\cdots,3}=-\frac{3}{2}B^1_{113,\cdots,3}=
    -3B^1_{123,\cdots,3}=\sqrt{\frac{3}{k(k+2)}} \label{Xi2}   
\end{eqnarray}
 For this definition of baryon states, the hypercharge extended to arbitrary
  $N_c$  is given by
\begin{equation}
Y= \frac{N_c B}{3} + S,
\end{equation}
where $B$ is the baryon number and $S$ is the 
strangeness which reduces to $Y=B+S$
in the physical case with $N_c=3$.
 
 We use the wave function (\ref{N}) of proton and
  neutron also for the case of flavor $SU(2)$ symmetry,
  where the subscript 3 means antisymmetric pair of $u$ and $d$
  quarks.   
 \par
 By use of the above baryon states
  we calculate the  matrix elements  
$<B_f| O^{a i}| B_i>$ and
 $<B_f | \lambda^a \sigma^i | B_i>$ where $a=0,~1,\cdots,~8$ and 
 $i=0,~1,~2,~3$.  We denote by $\lambda^0$  the unit matrix of flavor
  $SU(3)$ and
   $\sigma^0$ is the unit matrix of spin space representing spin-nonflip.
 These matrix elements correspond to the various physical
  quantities of baryons as the mass, the magnetic moment,
  the axial coupling constant etc.
\begin{description}
 \item[(1)]
  $<B \mid \lambda^0 \sigma^0 \mid B>$  $\cdots$
 the mass~~$m_B$
 \item[(2)]
 $<B \mid Q \sigma^i \mid B>$
 with $Q=\lambda^3/2 + \lambda^8/2\sqrt{3}$ $\cdots$
 the magnetic moment~~$\mu_B$,
 \item[(3)]
 $<B \mid \lambda^a \sigma^i \mid B>$
  $\cdots$
 the axial vector coupling constant~~$g_A$,
 \item[(4)]
 $<B_f \mid \lambda^{4+i5} \sigma^i \mid B_i>$
  $\cdots$
 where $SU(3)$ matrix $\lambda^{4+i5}$ changes $s$ quark to $u$ quark
 and corresponds to raising operator of the $V$-spin $V^+$ and
 contributes to the semi-leptonic decay of hyperons.
 \item[(5)]
 $<B_f \mid \lambda^{6+i7} \sigma^0 \mid B_i>$
 $\cdots$
 where $SU(3)$ matrix $\lambda^{6+i7}$ changes $s$ quark 
 to $d$-quark and corresponds to the raising operator of the $U$-spin,
  $U^+$ and contributes to the non-leptonic decay of hyperons.
\end{description}
 
 In the case of the $SU(3)$ symmetry the most general flavor octet
  vertex constructed from spin 
 1/2 baryon is represented as a sum of two independent terms as follows: 
\begin{eqnarray}
<B_f \mid O^{a} \mid B_i> &=&
 {\cal M}{\rm tr}({\bar B_f}\lambda^a B_i) + {\cal N}{\rm tr}
({\bar B_f}B_i \lambda^a) \nonumber \\
&=& F{\rm tr}(\lambda^a [{\bar B}_f,B_i]) + D{\rm tr}
(\lambda^a \{{\bar B}_f,B_i\}), \label{FD}
\end{eqnarray}
for the baryon vertex, 
where $\lambda^a$ is  a flavor octet matrix with $a=1,...,$ 8.
This $F/D$ ratio will be denoted by $F_+/D_+$. 
For the spin-flip vertex $<B_f|O^{ai}|B_i>$ we have a similar expression
and the $F/D$ ratio is denoted by $F_-/D_-$.  
The $F/D$ ratios are useful to observe the large $N_c$-dependence of
QCD independently of details of the specific effective model of QCD.
 \par
The operators $O^{a0}$ and $O^{ai}$ with $i=1,$ 2, 3
 transform as $({\bf 8,1})$ and ${\bf(8,3)}$, respectively under
 $(SU(3)_f,SU(2)_s)$ symmetric transformations.
 \par

In the case of the $SU(2)$ flavor symmetry we also introduce the concept
of $F/D$ ratios given by (\ref{FD}) as in the case of $SU(3)$.
 Furthermore we assume that the isospin-nonflip baryon vertex 
  corresponds to the generator $\lambda^8$ of the $SU(3)$.  
\par 
\section{The $F/D$ Ratios from $SU(4)$ and $SU(6)$ NRQMs}
 In the NRQMs with the $SU(4)$ and $SU(6)$ symmetries
 the spin 1/2 baryons are given by the completely symmetric
 representation with respect to spin and flavor which are represented
 by the Young diagrams with the first row of length $k+1$  and the 
 second row of length $k$ where $k$ is given by $N_c =2k+1$
 ($k=0,1,2,\cdots)$. Their dimensions are  $\bf {}_6 H_{N_c}$ and
 $\bf {}_4 H_{N_c}$ for $SU(4)$ and $SU(6)$ symmetry,
  where  ${}_n H_r$ is a repeated 
 combination ${}_n H_r = {}_{n+r-1} C_r = (n+r-1)!/r!(n-1)!$.
 \par
  In order to calculate the $F/D$ ratio of spin-nonflip baryon vertices
  in the case of two flavors, we consider the vertex which transform
  as the charge operator $Q=(\lambda^3 + \lambda^8/\sqrt{3})/2$
  under the $SU(3)$ transformation and reduce to  $Q=(\tau^3 +1)/2$
  for arbitrary $N_c$.
  \par
 The vertex is given by
\begin{eqnarray}
& & Q_p = Q_F + \frac{1}{3} Q_D,\label{fdnp} \\
& & Q_n  =  - \frac{2}{3} Q_D,\label{fdnn}
\end{eqnarray}
where $Q_F$ and $Q_D$ are the $F$ and $D$ type contributions to the 
operator ${Q}$.
 In the case of the $SU(4)$ symmetric NRQM with two flavors
  the isospin-flip or isovector  and isospin-nonflip or
  isoscalar baryon vertices are given by
\begin{eqnarray}
& & Q_{SU(4)}^{I=1} = \left(
 \begin{array}{ccc}
  {\bf 15}&{\bf  {}_4 H_{N_c}}&{\bf {}_4 H_{N_c}}^* \\
  \lambda^3 \sigma^0 & B & {\bar B}
 \end{array}
 \right)c, \label{Q1} \\
& & Q_{SU(4)}^{I=0} = 
 \left(
 \begin{array}{ccc}
  {\bf 15}&{\bf  {}_4 H_{N_c}}&{\bf {}_4 H_{N_c}}^* \\
  \sigma^0 & B & {\bar B}
 \end{array}
 \right)c, \label{Q0}
 \end{eqnarray}
 where $c$ is an unknown constant in the leading order of $1/N_c$.
 \par
 It is noted here that in the $SU(4)$ symmetric NRQM, both of the  isoscalar
 part and isovector part of the spin-flip vertex are generators which 
 belong to the same $ SU(4)$ supermultiplet {\bf15}. 
  \par
 In the $SU(3)$ case we consider the $S$-wave non-leptonic hyperon decay to
 study the  spin-nonflip baryon vertex
  $<B_f | \lambda^{4+i5} \sigma^0 |B_i >$.
  In general the effective Hamiltonian of hyperon decay
 transforms as an element of($SU(3)_L \times SU(3)_R$) =  ({\bf 8,1})
  and ({\bf 27,1}).  In this paper
 we assume  the octet dominance or $\Delta I = 1/2$ 
 enhancement$\cite{Bardeen}$.
 \par
 The baryon vertex of the non-leptonic decay $B_i \to B_f + \pi^-$ is
  given by  \begin{eqnarray}
{\cal V}(B_i \to B_f + \pi^-) 
&=& <B_f \mid [I^+,U^+] \mid B_i>  \nonumber\\
&=& <B_f \mid V^+ \mid B_i>, 
\end{eqnarray}
where $I^+ = I^1 + iI^2$, $U^+ = U^1 + iU^2$ and $V^+ = V^1 +iV^2$ are
the raising operators of isospin, $U$-spin and $V$-spin, respectively
 in the $SU(3)$ flavor
space. In terms of the $F_+/D_+$ ratio defined by (\ref{FD}) the 
$S$-wave decay vertices
of $\Sigma^- \to n + \pi^-$ and $\Xi^- \to \Lambda + \pi^-$  for arbitrary 
$N_c$ are expressed as
\begin{eqnarray}
& & {\cal V}(\Sigma^- \to n + \pi^-) =
 \frac{1}{k}(F_+ -D_+),\label{fdp} \\
& & {\cal V}(\Xi^- \to \Lambda + \pi^-) =
 \frac{1}{k+2}\sqrt{\frac{3}{2k}}\{(2k+1)F_+ -D_+\}.\label{fdn}
\end{eqnarray}
 In the $SU(6)$ NRQM, the baryon vertex of the non-leptonic hyperon decay 
 is given by
\begin{equation}
 {\cal V}(B_i \to B_f + \pi^-)_{SU(6)} = 
 \left(
 \begin{array}{ccc}
  {\bf 35}& {\bf {}_6 H_{N_c}}&{\bf {}_6 H_{N_c}}^* \\
  \lambda^{(4+i5)}\sigma^0 & B & {\bar B}
 \end{array}
 \right) c'.
\end{equation}
 Here the isoscalar part and isovector part of the spin-nonflip vertex
 belong to the same supermultiplet {\bf 35} of $SU(6)$.
\par
In the $SU(6)$ NRQM the $S$-wave decay vertices
of $\Sigma^- \to n + \pi^-$ and $\Xi^- \to \Lambda + \pi^-$  for arbitrary 
$N_c$ are  given by
\begin{eqnarray}
& & {\cal V}(\Sigma^- \to n + \pi^-)_{SU(6)} = 2c', \label{fdp1} \\
& & {\cal V}(\Xi^- \to \Lambda + \pi^-)_{SU(6)} = \sqrt{6k}c', \label{fdn1}
\end{eqnarray}
where $c'$ is an unknown constant in the leading order of $1/N_c$.
\par
 In the $SU(2)$ case we can obtain the $F/D$ ratio of the spin-nonflip
  baryon
 vertex for the ``baryon" with spin 1/2 of arbitrary $N_c$ by comparing
 (\ref{fdnp}) and (\ref{fdnn}) with (\ref{Q1}) and (\ref{Q0}).
 In the $SU(3)$ case, we obtain the same $F/D$ ratio from (\ref{fdp}),
 (\ref{fdn}), (\ref{fdp1}) and (\ref{fdn1}). The obtained result is
\begin{eqnarray}
\left(\frac{F_+}{D_+}\right)_{SU(4),SU(6)} =
 -\frac{N_c + 1}{N_c - 3}. 
\end{eqnarray}
\hspace*{\parindent}
The same value of $F/D$ ratio for the $SU(4)$ NRQM with that of
the $SU(6)$ model is due to the fact that the wave functions for nucleons
are the same in $SU(4)$ and $SU(6)$ NRQM contrary to the case of CSM.
 The same result can  also be derived by the algebraic method.
 It is noted that in the case of NRQMs with
 $SU(4)$ and $SU(6)$ symmetries the amplitude of non-leptonic hyperon 
 decay depends on the way of extrapolation to the 
 large $N_c$ baryon. However the $F/D$ ratio does not depend on the way of
 extrapolation to large $N_c$ baryons. 
 \par
 In the $SU(4)$ and $SU(6)$ NRQMs
 the $F_+/D_+$ ratio tends to $-1$
 in the limit of $N_c \to \infty$. For $N_c=1$ the $F_+/D_+$ ratio becomes 
 $1$ reflecting the fact that the baryons are quarks themselves.
For $N_c=3$ the ratio  becomes $\infty$ reflecting the fact that the quark
 number is conserved implying the pure $F$-type amplitude$\cite{Sakita}$.
 \par
 Next we turn to the $F_-/D_-$ ratio of the spin-flip baryon vertex  
  for arbitrary $N_c$ $\cite{Takamura}\cite{Karl}$.
 In the $SU(4)$ NRQM the magnetic moment of baryon $B$ is given by
 \begin{equation}
(\mu_B)_{SU(4)} =
 \left(
 \begin{array}{ccc}
  {\bf 15}&{\bf  {}_4 H_{N_c}}&{\bf {}_4 H_{N_c}}^* \\
  \sigma^3 Q& B& {\bar B}
 \end{array}
 \right) \mu
\end{equation}
 where $\sigma^3$ represents spin up or down and $Q =( \tau^3 + 1)/2$ the
 charge of the nucleon and $\mu$ is an unknown constant to the 
 leading order in $1/N_c$.
 It is noted here that in the $SU(4)$ NRQM, both of the isoscalar part
 and isovector part of the spin-flip vertex are generators which belong
 to the same $SU(4)$ supermultiplet {\bf15}.
 \par
 Similarly, in the $SU(6)$ NRQM case  the magnetic moment of spin
 1/2 baryon $B$ is given by
\begin{equation}
 (\mu_B)_{SU(6)} =
 \left(
 \begin{array}{ccc}
  {\bf 35}& {\bf {}_6 H_{N_c}}&{\bf {}_6 H_{N_c}}^* \\
  \sigma^3 Q& B& {\bar B}
 \end{array}
 \right) \mu,
\end{equation}
where $Q=\lambda^3 /2+ \lambda^8/2\sqrt{3}$ is the charge operator.
The isoscalar part and isovector part of the spin-flip vertex
 belong to the same supermultiplet {\bf 35} of $SU(6)$.
 \par
 Both of the $SU(4)$ and $SU(6)$ NRQMs  give the same magnetic
 moment of nucleons for arbitrary $N_c$:
\begin{eqnarray}
& & (\mu_p)_{SU(4),SU(6)} = (k+2)\mu,\label{mup} \\
& & (\mu_n)_{SU(4),SU(6)} = -(k+1)\mu.\label{mun}
\end{eqnarray}
In terms of the  $F/D$ ratio the magnetic moments of nucleons are given by
\begin{eqnarray}
& & \mu_p = \mu_F + \frac{1}{3} \mu_D, \label{fdpm}\\
& & \mu_n = -\frac{2}{3} \mu_D ,\label{fdnm}
\end{eqnarray}
where $\mu_D$ and $\mu_F$ are the $D$ and $F$ type contributions to the 
 magnetic moment of baryons.
  Comparing  (\ref{mup}) and (\ref{mun}) with (\ref{fdpm}) and (\ref{fdnm})
 we obtain the
 $F/D$ ratio of spin-flip baryon vertex  for the ``baryon" with spin 1/2  
 and arbitrary $N_c$. The obtained result is
\begin{eqnarray}
 \left(\frac{F_-}{D_-}\right)_{SU(4),SU(6)}=\frac{N_c + 5}{3(N_c + 1)}.
\end{eqnarray}
\hspace*{\parindent}
 In the $SU(4)$ and $SU(6)$ NRQMs
 the $F_-/D_-$ ratio tends to 1/3
 in the limit of $N_c \to \infty$. For $N_c =1$ the $F/D$ ratio  becomes 1 
 reflecting the fact that the baryons are quarks themselves.
 For physical baryon with $N_c=3$ the $F/D$ ratio takes 
 the familiar value 2/3 of the $SU(6)$ NRQM$\cite{Sakita}$.

\par
Here we point out that there is a relation between two $F/D$ ratios
of spin-nonflip and spin-flip baryon vertices in the NRQM$\cite{Matsuoka}$
independent of $N_c$;
\begin{equation}
(4 f_+ + 1)(4 f_- -1 ) =3,\label{fdrelation}
\end{equation}
where $f_{\pm} = F_{\pm}/(F_{\pm} + D_{\pm})$.


\section{The $F/D$ Ratios in $SU(2)$ and $SU(3)$ CSMs}
\hspace*{\parindent}
 In the $SU(2)$ CSM the spin 1/2 baryon state is represented
 by the elements of $SU(2)$ matrix in the fundamental representation of 
 $SU(2)$ which is independent of color degrees of freedom $N_c$.
 \par
 In $SU(2)$ CSM the isoscalar part and the isovector part of currents for 
 the baryon
 have distinct origins. That is, the isovector part is the space 
 integral of the time component of 
 conserved isovector current which is the Noether current
 reflecting the symmetry of the chiral Lagrangian and is
given by 
\begin{equation}
  J^\mu
= f_\pi^2{\rm tr}[(\partial_\mu U U^\dagger)\Omega]
+ \frac{i}{8e^2}{\rm tr}\{[\partial_\nu U U^\dagger,\Omega]
 [\partial^\mu U^\dagger,\partial^\nu U U^\dagger]\},\label{Jm}
\end{equation}
where $\Omega$ is the generator of $SU(2)$ flavor symmetry.
\par
On the other hand the isoscalar part 
 comes from the space integral of the time component
 of the  baryon number current which is topologically conserved;
\begin{equation}
  B^{\mu}
= \frac{\epsilon^{\mu\nu\alpha\beta}}{24\pi^2}{\rm tr}
  [(U^\dagger \partial_\nu U)(U^\dagger \partial_\alpha U)
  (U^\dagger \partial_\beta U)].\label{Bm}
\end{equation}
Thus there is no direct relation between the isovector current $J^{\mu}$
 and the isoscalar part $B^{\mu}$$\cite{Adkins}$.
  \par
  In order to calculate the $F/D$ ratio of spin-nonflip baryon vertices
  we consider the charge operator 
 $Q$  of  nucleon. The vertex is given by
\begin{equation}
  Q_{SU(2) \; CSM} = 
  \left(
  \begin{array}{ccc}
  {\bf 1}& {\bf 2} & {\bf 2} \\
   \tau^0 & B& {\bar B}
  \end{array}
  \right)
  \left(
  \begin{array}{ccc}
   {\bf 1}& {\bf 2}& {\bf 2} \\
   \sigma^0& B& {\bar B}
  \end{array}
  \right)c +\cdots. \label{mub}
\end{equation}
 The ellipsis in (\ref{mub}) denotes contributions from time
 derivative of dynamical variables of the spin-isospin rotation
 of the chiral soliton where the isoscalar part of magnetic moment
 given by the topological or baryon number current is contained. 
\begin{eqnarray}
& & Q_{p, \, SU(2) \; CSM} = c
 + \cdots,  \\
& & Q_{n, \, SU(2) \; CSM} = c
 + \cdots.
\end{eqnarray}
 Thus the $F/D$ ratio in the $SU(2)$ CSM is
\begin{equation}
\left(\frac{F_+}{D_+}\right)_{SU(2) \; CSM}= -1 +\cdots,
\end{equation}
\hspace*{\parindent}
 On the other hand in the NRQM the isovector and isoscalar parts
  of the spin-flip
  baryon vertex are space integrals of space component of
 Noether current and  topological current, respectively.
  \par
 Therefore the magnetic moment of the spin 1/2 baryon is given by
\begin{equation}
 (\mu_B)_{SU(2) \; CSM} = 
  \left(
  \begin{array}{ccc}
  {\bf 3}& {\bf 2} & {\bf 2} \\
   \tau^3 & B& {\bar B}
  \end{array}
  \right)
  \left(
  \begin{array}{ccc}
   {\bf 3}& {\bf 2}& {\bf 2} \\
   \sigma^3& B& {\bar B}
  \end{array}
  \right) \mu^{I=1}  +\cdots, \label{mubm}
\end{equation}
where $\tau^3 /2 = Q^{I=1}$ is the isovector part of charge. The 
 ellipsis in (\ref{mubm}) denotes contributions from the time derivative
 of dynamical variables where the isoscalar part are contained.
  The magnetic moments of the nucleons are 
\begin{eqnarray}
& & (\mu_p )_{SU(2) \; CSM} = \frac{1}{2}\mu_{I=1} + \cdots,  \\
& & (\mu_n )_{SU(2) \; CSM} = -\frac{1}{2}\mu_{I=1} + \cdots.
\end{eqnarray}
{}From these results and (\ref{fdpm}) and (\ref{fdnm})
we obtain the $F_-/D_-$ ratio of spin-flip baryon vertex
 in the $SU(2)$ CSM as
\begin{equation}
\left(\frac{F_-}{D_-}\right)_{SU(2) \; CSM}= \frac{1}{3} +\cdots.
\end{equation}
\hspace*{\parindent}
 In the $SU(3)$ CSM the spin 1/2 baryon 
 states are represented by the $SU(3)$ matrix elements of the
 ${\bf (1,k)}={\bf (1+k)(3+k)}$ dimensional representation 
 where $k=(N_c -1)/2$. ${\bf (1,k)}$ denotes the representation
 of $SU(3)$ with the Young diagram
 which  has the first row of length $k+1$ and the second row of length $k$.
 In the case $N_c=3$ this is the octet or the regular representation of
  $SU(3)$. \par 
The baryon vertex of $S$-wave non-leptonic hyperon decay 
in the $SU(3)$ CSM is given by
\begin{eqnarray}
& & {\cal V}(B_i \to B_f + \pi^-)_{SU(3) \; CSM} \nonumber \\
&=& \sum_n
  \left(
  \begin{array}{ccc}
  {\bf 8}& {\bf (1,k)}& {\bf (1,k)}_n^* \\
  \lambda^{4+i5} & B& {\bar B}
  \end{array}
  \right)
  \left(
  \begin{array}{ccc}
  {\bf  8}&{\bf  (1,k)}& {\bf (1,k)}_n^* \\
  \sigma^0& B& {\bar B}
  \end{array}
  \right)c' + \cdots,
\end{eqnarray}
where the summation over $n$ means two orthogonal states of baryons of 
 $\bf (1,k)^*$ representation and $c'$ is a constant. 
 The ellipsis denote  corrections from
 the time derivative of the $SU(3)$ matrix valued dynamical variable
 $A(t)$ describing the ``rotations" in spin-flavor space and contains
 the higher order term of $1/N_c$ expansion.
 \par
 The $S$-wave decay vertices
of $\Sigma^- \to n + \pi^-$ and $\Xi^- \to \Lambda + \pi^-$  for arbitrary 
$N_c$ are expressed as
\begin{eqnarray}
& & {\cal V}(\Sigma^- \to n + \pi^-)_{SU(3) \; CSM} = 
2(k+3)c' + \cdots, \\
& & {\cal V}(\Xi^- \to \Lambda + \pi^-)_{SU(3) \; CSM} = 
\sqrt{6k}c' + \cdots.
\end{eqnarray}
 {}From these results and (\ref{fdnp}) and (\ref{fdnn})
  we obtain the $F_+/D_+$ ratio of spin-nonflip baryon
 vertex in the $SU(3)$ CSM
\begin{eqnarray}
\left(\frac{F_+}{D_+}\right)_{SU(3) \; CSM}
= - \frac{N_c^2 + 4 N_c - 1}{N_c^2 + 4 N_c - 9} +\cdots.
\end{eqnarray}
 The $F_+/D_+$ ratio tends to $-1$ in the limit $N_c \to \infty$ 
 and it is 1 and $-5/3$ for $N_c =1$  and  $N_c=3$,
 respectively$\cite{Nappi}$$\cite{Bijinen}$.
 \par
 In Fig.~2 we compare the $D_+/F_+$ ratios in the $SU(6)$ NRQM
 and the ratio in the $SU(3)$ CSM.
Here $D_+/F_+$ ratios are shown instead of $F_+/D_+$
in order to avoid singular behaviors of the $F_+/D_+$ which appear 
at $N_c=3$ and $-2+\sqrt{13}$ in the NRQM and CSM, respectively. 
 Two ratios coincide at $N_c = 1$ and $N_c \to \infty$  tending to $-1$.
  The experimental value of the 
 $F/D$ ratio for spin-nonflip baryon vertex $(F_+/D_+)_{\rm exp}
  = -3.0 \pm 0.2$ or $(D_+/F_+)_{\rm exp} = -0.33 \pm 0.02$
 lies also
 between those of NRQM and CSM.

\par

Next we turn to the $F_-/D_-$ ratio of spin-flip vertex.
 In  the $SU(3)$ chiral soliton model
 differently from the $SU(2)$ CSM, 
 the magnetic moment of spin 1/2 baryon $B$ is given by
 $\cite{Takamura}\cite{Patera}$
\begin{equation}
 (\mu_B)_{SU(3) \; CSM} = \sum_n
  \left(
  \begin{array}{ccc}
  {\bf 8}& {\bf (1,k)}& {\bf (1,k)}_n^* \\
   Q& B& {\bar B}
  \end{array}
  \right)
  \left(
  \begin{array}{ccc}
  {\bf  8}&{\bf  (1,k)}& {\bf (1,k)}_n^* \\
   \sigma^3& B& {\bar B}
  \end{array}
  \right) \mu +\cdots
\end{equation}
where the summation over $n$ means two orthogonal states of baryons of 
 $\bf (1,k)$ representation and the ellipsis denote  corrections from
 the time derivative of the $SU(3)$ matrix valued dynamical variable
 $A(t)$ describing the ``rotations" in spin-flavor space which contain
 the higher order term of $1/N_c$ expansion.
 The results for the magnetic moments are
\begin{eqnarray}
& & (\mu_p)_{SU(3) \; CSM} = \frac{k+3}{3(k+4)} \mu  + \cdots, \\
& & (\mu_n)_{SU(3) \; CSM} = \frac{k^2+5k+3}{3(k+2)(k+4)} \mu + \cdots.
\end{eqnarray}
 There is a difference in the magnetic moments of nucleons in the chiral
 soliton models between flavor 2 and 3 contrary to the NRQM. This comes
 from the fact that the nucleon states belong to the fundamental
 representation ${\bf 2}$ in the $SU(2)$ soliton irrespective of $N_c$ 
 while in the $SU(3)$ case the states belong to the regular
 representation ${\bf (1,k)}$.
 \par
  {}From these results we obtain the $F/D$ ratio of spin-flip baryon vertex
 in the $SU(3)$ CSM
\begin{eqnarray}
\left(\frac{F_-}{D_-}\right)_{SU(3) \; CSM}
=\frac{N_c^2 + 8 N_c + 27}{3(N_c^2 + 8N_c + 3)} +\cdots.
\end{eqnarray}
 In the $SU(2)$ and $SU(3)$ CSMs the $F/D$ ratio becomes $1/3$ if we take
 the limit $N_c \to \infty$. For $N_c=1$, the $F/D$ ratio becomes $1$ and
 for $N_c=3$ it is $5/9$$\cite{Sakita}$.
 \par
 In Fig.~3 we compare the $F/D$ ratios of spin-flip baryon vertex in the
 $SU(6)$ NRQM
 and in the $SU(3)$ CSM.
 Two ratios coincide at $N_c = 1$ and $N_c \to \infty$. The experimental
 value of the 
 $F/D$ ratio for  spin-flip baryon vertex $(F_-/D_-)_{\rm exp}
 = 0.58 \pm 0.04$ lies between the two lines nearer
 to that of $SU(3)$ chiral soliton.
 
 \par
 In the case of CSM the relation between $F/D$ ratios of spin-nonflip and
 spin-flip
 baryon vertices given by (\ref{fdrelation}) which holds in the NRQM is not
 satisfied.
 
 
\section{The $1/N_c$ Expansion of $F/D$ Ratios in the NRQM and the CSM 
and the $I=J$ rule}
\hspace*{\parindent}
 {}From the previous calculation, we find the value of $F_{\pm}/D_{\pm}$
 ratios 
 of both spin-flip and nonflip baryon vertices for arbitrary
  $N_c$. In order to investigate the physical meaning of the obtained 
  $F/D$ ratios of baryon
 vertices, we expand the $F/D$ ratios assuming $N_c$ is large 
  as follows,\begin{eqnarray}
& & \left(\frac{F_+}{D_+}\right)_{SU(4),SU(6)}
 =  -\frac{N_c+1}{N_c-3} 
 =  -1 - \frac{4}{N_c} + \frac{12}{N_c^2} + \cdots, \label{fdnrqm} \\
& & \left(\frac{F_+}{D_+}\right)_{SU(3) CSM}
 =  -\frac{N_c^2+4N_c-1}{N_c^2+4N_c-9}
 =  -1 + \frac{0}{N_c} - \frac{8}{N_c^2} + \cdots, \label{fdcsm} \\
& & \left(\frac{F_+}{D_+}\right)_{SU(2) CSM}
 =  -1 + \cdots.
\end{eqnarray}
Similarly,
\begin{eqnarray}
& & \left(\frac{F_-}{D_-}\right)_{SU(4),SU(6)}
 =  \frac{N_c+5}{3(N_c+1)}
 =  \frac{1}{3} + \frac{4}{3N_c} - \frac{4}{3N_c^2} + \cdots,
 \label{fdnrqm1} \\
& & \left(\frac{F_-}{D_-}\right)_{SU(3) CSM}
 =  \frac{N_c^2 + 8N_c +27}{3(N_c^2 + 8N_c +3)}
 =  \frac{1}{3} + \frac{0}{N_c} + \frac{8}{N_c^2} + \cdots,
\label{fdcsm1} \\
& & \left(\frac{F_-}{D_-}\right)_{SU(2) CSM}
 =  \frac{1}{3} +\cdots.
\end{eqnarray}
\hspace*{\parindent}

The $1/{N_c}$ expansions of $F_+/D_+$
for both the NRQM and CSM do not converge at $N_c= 3$. The expansion 
 ($\ref{fdnrqm}$) in the NRQM converges for $N_c > 3$ while
the expansion (\ref{fdcsm}) in the CSM does at $N_c>2+\sqrt{13}=5.6\cdots$.
On the other hand the expansion ($\ref{fdnrqm1}$) in the NRQM
 converges for $N_c >1$, but the expansion ($\ref{fdcsm1}$) in the CSM
  converges at $N_c>4+\sqrt{13}=7.6\cdots$. 
 \par
 First we consider the physical meaning of the limiting value $-1$
 and $1/3$ for the $F/D$ ratios and introduce a ``charge"
 which represents both for spin-flip and nonflip nucleon vertices
 with arbitrary $N_c$
 \begin{equation}
<B \mid {\tilde Q} \mid B> \equiv {\tilde \mu}_B,
\end{equation}
where $B$ is $p$ or $n$. By the use of wave functions of nucleon 
($\ref{N}$)
the isovector and isoscalar parts are given by
\begin{eqnarray}
& & {\tilde \mu}^{I=1} =
 {\tilde \mu}_p - {\tilde \mu}_n = (F+D)\mu  \\
& & {\tilde \mu}^{I=0} =
 {\tilde \mu}_p + {\tilde \mu}_n = (F-\frac{1}{3}D)\mu 
\end{eqnarray}
 In the case of spin-nonflip baryon vertex, if we substitute 
 the $1/N_c$ expansions of $F_+/D_+$  we find 
 ${\tilde \mu}^{I=1}$
 is $O(N_c^0)$ and ${\tilde \mu}^{I=0}$ is $O(N_c)$. On the other hand
 in the case of spin-flip baryon vertex, we obtain the result that
  ${\tilde \mu}^{I=1}$
 is $O(N_c)$ and ${\tilde \mu}^{I=0}$ is $O(N_c^0)$.
  Namely $I=J$ part of
 the baryon vertex is enhanced compared to $I \ne J$ part
 as is summarized in Table 1. 
 The $I=J$ enhancement is a consequence of the large $N_c$ counting
  rules for spin-nonflip and spin-flip baryon vertices.
\par

\begin{table}
\begin{center}
\begin{tabular}{|c|c|c|} \hline
 {\it   }& $I=0$ & $I=1$ \\ \hline
 $J=0$ & ${\tilde \mu}$($O(N_c)$) & ${\tilde \mu}$($O(N_c^0)$) \\ \hline
 $J=1$ & ${\tilde \mu}$($O(N_c^0)$) & ${\tilde \mu}$($O(N_c)$) \\ \hline
\end{tabular}  
\end{center} 
\caption{the $I=J$ rule for the charge operator in the chiral soliton model}
\end{table}
 Generally the isovector part  is suppressed by $1/N_c$ in comparison to
  the isoscalar part of the spin-nonflip vertex while the isoscalar part 
 is suppressed by $1/N_c$ in comparison to the isovector part of the
 spin-flip vertex.
 Therefore we can neglect the isoscalar part of spin-flip baryon vertex
 and the isovector part of spin-nonflip baryon vertex in the large $N_c$
 limit. Conversely  the limiting value $F/D=-1$ in the spin-nonflip
 vertex and $F/D=1/3$ in the spin-flip vertex are derived
  if we assume the $I=J$ enhancement.
 \par
The $I=J$ enhancement is seen in the low energy hadron phenomena.
 For instance the tensor coupling of $g_{\rho NN}$,
 which is spin-flip vertex, is larger than the
 vector coupling of $g_{\rho NN}$, which is spin-nonflip
 vertex, and the vector coupling of
 $g_{\omega NN}$ is larger than the tensor coupling of $g_{\omega NN}$
 as displayed in Table 2.
 This is a consequence of $1/N_c$ expansion$\cite{Mattice}$.

\begin{table}
\begin{center}
\begin{tabular}{|c|c|c|} \hline
 {\it   }& Vector & Tensor \\ \hline
 $g_{\rho \pi\pi}$ & $O(1/\sqrt{N_c})$ & $O(\sqrt{N_c})$ \\ \hline
 $g_{\omega \pi\pi}$ & $O(\sqrt{N_c})$ & $O(1/\sqrt{N_c})$ \\ \hline
\end{tabular}  
\end{center} 
\caption{the $I=J$ rule for $g_{\rho \pi\pi}$ and $g_{\omega \pi\pi}$ coupling
 constant}
\end{table}

 In the $SU(2)$ CSM this $I=J$ rule is explained by the fact that
 the isovector part of spin-nonflip baryon
 vertex comes from the time component of
 the Noether current which is $O(N_c^0)$ and the isoscalar part comes from 
 that of the 
 topological current which is $O(N_c)$. On the other
 hand the isovector part of spin-flip baryon vertex comes from the space
 component of Noether current $J^{5 \mu}$
 which is $O(N_c)$ and the isoscalar part comes from that of 
  topological current which is
 $O(N_c^0)$(See Table 3).
 
\begin{table}
\begin{center}
\begin{tabular}{|c|c|c|} \hline
 {\it   }& $I=0$ & $I=1$ \\ \hline
 $J=0$ & $B^0$($O(N_c)$) & $V^0$($O(N_c^0)$) \\ \hline
 $J=1$ & $B^i$($O(N_c^0)$) & $V^i$($O(N_c)$) \\ \hline
\end{tabular}  
\end{center} 
\caption{the $I=J$ rule for the charge operator}
\end{table}

 \par
 We  note here  that the $1/N_c$ correction to $F/D$
 ratio has two different origins. One is the $1/N_c$ correction from the
 baryon  wave function
  and the other is from the $1/N_c$ correction of the dynamical
 quantities. In the $SU(4)$ symmetric model the $1/N_c$ correction comes
 only from the baryon wave functions and not from the vertex
  operators. Contrary 
 to the $SU(4)$ symmetric NRQM, in the $SU(2)$ CSM
 the $1/N_c$ corrections come from the dynamical variables
 (time derivative of spin-isospin rotation matrix $A(t)$) not from the
 baryon wave functions.
 \par
Next we comment on  the $1/N_c$ correction in the $SU(3)$ CSM.
 We calculated
 the $F/D$ ratios from the nonleptonic hyperon decay and the magnetic moment.
 There are, however, 
  additional contributions to the nonleptonic hyperon decay vertices
 and the magnetic moments from  the correction due to the
  soliton rotation in the spin-isospin space.
By taking these effects into account, we obtain the additional contributions
 to the $F/D$ ratios which are suppressed by $1/N_c$.
  Therefore there exist the $1/N_c$ corrections
 even if we consider the $SU(3)$ CSM.
\par
 Let's consider the reason why the $I=J$ rule works in the $SU(2)$ CSM.
 In the $SU(2)$ CSM the spin and the isospin operator are
\begin{eqnarray}
& & I^a = i\lambda {\rm tr}(A^\dagger \tau^a {\dot A}),\label{Ia} \\
& & J^i = -i\lambda {\rm tr}(\sigma^i A^\dagger {\dot A}),\label{Ji}
\end{eqnarray}
where $\lambda$ is a moment of inertia of the order $O(N_c)$ and
 ${\dot A}$ is a time
 derivative of $A$.
 \par
 If spin and isospin of baryon state are of $O(N_c^0)$, then from ($\ref{Ia}$) 
 and ($\ref{Ji}$) we find that  ${\dot A}$ is
 $O(1/N_c)$. The space component of Noether current ($\ref{Jm}$) and
  the time component
 of baryon number current ($\ref{Bm}$) are suppressed compared
  to the time component of
 Noether current and the space component of baryon number current i.e.
the  $I=J$ rule($I \ne J$ suppression) is satisfied.
 \par
 But if spin and isospin of baryon state are of $O(N_c)$, then ${\dot A}$ is
 $O(N_c^0)$. The space component of Noether current and the time component
 of baryon number current are {\it not} suppressed compared to the time
 component of Noether current and the space component of baryon number
 current i.e. the $I=J$ rule breaks down($I \ne J$ enhancement).
 \par
 It is desirable of course we need to understand the $I=J$ rule from  quarks
 and gluons i.e. from QCD.
 
  
\section{The model independent analysis of the F/D ratios in the
limit $N_c \to
 \infty$}
 In this section we study the physical meaning of the $F/D$ ratios
  in the limit  $N_c \to
 \infty$ in more detail. The matrix element of the diagonal operator $H^8$ 
  can be expressed in general  according to  the Wigner-Eckart
 theorem$\cite{Nachatoman}$ as follows;
\begin{eqnarray}
& & <B_f \mid H^8 \mid B_i> \nonumber \\
&=& a \; Y + b \big\{ I(I+1)-\frac{Y^2}{4}
-\frac{{\bf F}^a {\bf F}^a}{3} \bigr\},
\end{eqnarray}
where ${\bf F}^a$ is the $SU(3)$ generator and ${\bf F}^a {\bf F}^a$
 is the  Casimir operator of the $SU(3)$ and
\begin{equation}
{\bf F}^a {\bf F}^a = \frac{(k+2)^2}{3},
\end{equation}
for the ${\bf (1,k)}$ representation of $SU(3)$.  
 \par
 On the other hand we can calculate the matrix element of the diagonal
 operator $H^8$ using $F$ and $D$ and baryon states given by $(\ref{N})
  \sim (\ref{Xi2})$.
\begin{eqnarray}
& & <N \mid H^8 \mid N > = -F+\frac{1}{3}D \label{Nfd}\\
& & <\Sigma \mid H^8 \mid \Sigma > =
 -F+\frac{1}{3}D + \frac{1}{k}(F-D) \label{Sigmafd}\\
& & < \Lambda \mid H^8 \mid \Lambda > =
 -F+\frac{1}{3}D + \frac{1}{k}(F-D) + \frac{2}{k+2}
\{F+D - \frac{1}{k}(F-D)\} \label{Lambdafd}\\
& & <\Xi \mid H^8 \mid \Xi > 
 = -F+\frac{1}{3}D + \frac{2}{k}(F-D) + \frac{3}{k+2}
\{F+D - \frac{1}{k}(F-D)\} \label{Xifd}
\end{eqnarray}
The general expression that satisfy these equations is
\begin{eqnarray}
& & <B_f \mid H^8 \mid B_i> \nonumber \\
&=& -F + \frac{1}{3}D + \frac{2}{k}(F-D) K \nonumber \\
&+& \frac{1}{k+2}\{F+D - \frac{1}{k}(F-D)\} \bigl\{ 
{ I(I+1)-(K+\frac{1}{2})
(K+\frac{3}{2})}\bigr\}, \label{Gfd}
\end{eqnarray}
where ${ K=-S/2}$, $S$ being the strangeness.
 \par
 {}From (\ref{Nfd}) $\sim$ (\ref{Xifd}) and ($\ref{Gfd}$) we find the following
 structure among the matrix elements of flavor $SU(3)$ generator $\lambda^8$.
 The first part of (\ref{Gfd})
 gives the same contribution  $-F + D/3$
 to all states. The second part proportional to $F-D$ which is 
 of the order $O(1/N_c)$  gives increasing
 contributions with $K$.
The third part which includes both $O(1/N_c)$ term proportional
 to $F+D$ and $O(1/N_c^2)$ term proportional
 to $F-D$ appears only for the states located on the inner triangle
 of the root diagram because of the factor $\{ I(I+1) - (K+1/2)(K+3/2)\}$. 
 \par   
Using the definition  $Y=N_cB/3 + S$, we can rewrite the above expression as 
\begin{eqnarray}
& & <B_f \mid H^8 \mid B_i> \nonumber \\
&=& -\frac{1}{6k(k+2)}\{(k+5)(2k+1)F+(k-1)(2k+5)D\} Y \nonumber \\
& & - \frac{1}{k+2}\bigl\{ F+D-\frac{1}{k}(F-D)\bigr\}
  \bigl\{ I(I+1)-\frac{Y^2}{4}-\frac{(k+2)^2}{9}\bigr\}.
\end{eqnarray}
 Here we note that
  we {\it have to} define the hypercharge as $Y=N_c B/3 + S$ which is
 consistent
  with the extended baryon states expressed by (\ref{N}) $\sim$ (\ref{Xi2}).
 The strangeness
  $S$ and the baryon number $B$ are the quantities of order $O(1)$, thus the
  extended hypercharge $Y$ is of order $O(N_c$) so that
 the center of the root diagram for  baryon multiplet in 
  the ${\bf (1,k)}$ 
 representation is located at the  origin of the $(I^3,Y)$ plane.
  
 On the other hand 
if we define the hypercharge as $Y=B+S$, we cannot obtain the result that is
 consistent with the Wigner-Eckart Theorem.
 \par
 The generator ${\bf F^a}$ of $SU(3)$ flavor symmetry 
 is represented in terms of $(1+k)(3+k) \times (1+k)(3+k)$ matrix,
 the dimension of spin 1/2 baryon. 
 The general expressions of these  matrix elements 
  contain the Casimir operator.
 \par
 If $H^8$ transforms as $(SU(3)_f,SU(2)_s)={\bf (8,1)}$, then $F_+/D_+
  = -1 + O(1/N_c)$.
With this $F_+/D_+$ ratio of the  mass formula for the  baryons,
 the mass difference
 of baryons  is given by
\begin{equation}
    \Delta m_B
 =  N_c a + b {K} + \frac{c}{N_c}
    \{{ I(I+1)-(K+\frac{1}{2})(K+\frac{3}{2})}\},
\end{equation}
where $a$, $b$ and $c$ are constants independent of $N_c$ and 
isospin. We find that the terms which depend on the isospin and
the strangeness are suppressed to the order $O(1/N_c)$. 
A similar formula is also derived in Ref.{\cite{Dashen}}  
 \par
 \par
 On the other hand if $H^8$ transforms as $(SU(3)_f,SU(2)_s)={\bf (8,3)}$,
 then $F_-/D_- = 1/3$. We can apply this formula to the isoscalar part of
 the magnetic moment. Then we obtain
\begin{equation}
   \mu_B^{I=0}
 = a' + b'  K
 + c'\bigl\{ I(I+1)-(K+\frac{1}{2})(K+\frac{3}{2}) \bigr\}
\end{equation}
where $a'$, $b'$ and $c'$ are  constants  independent of $N_c$.
 \par
 The  isospin and strangeness dependence survives in this formula even if
  we take
 the limit $N_c \to \infty$.
 \par
 The isospin and strangeness dependent terms are not necessarily
 to suppressed at the lowest order in $1/N_c$
  expansion$\cite{Dashen2}\cite{Luty}\cite{Carone}\cite{Jenkins}$.
 \par
 {}From $(\ref{Nfd}) \sim (\ref{Xifd})$, we can derive the model
 independent and $N_c$ independent relations.
 \par
 The model independent relations are the Okubo-Gell-Mann mass
 relation  
\begin{equation}
3\Lambda + \Sigma = 2(N + \Xi) , \label{og}
\end{equation}
and the refined Okubo-Gell-Mann relation$\cite{Sawada}$
\begin{equation}
3 \Lambda + \Sigma^+ + \Sigma^- - \Sigma^0 = p + n+ \Xi^0 + \Xi^- ,\label{rog}
\end{equation}
which is more accurately satisfied by experimental masses are $N_c$-dependent. 
  These relations are not  trivial, since the multiplet belonging to
 spin 1/2 baryon is no longer an octet representation of $SU(3)$.
 \par
  Actually the Coleman-Glashow mass relation
\begin{eqnarray}
\Sigma^+ - \Sigma^- = p -n + \Xi^0 - \Xi^-
\end{eqnarray}
holds only for $N_c = 3$.
  In the case of Coleman-Glashow mass relation the
 relevant operator is $ H^3$ which transforms as generator $\lambda^3$.
 The diagonal matrix elements of  operator $H^3$ for the octet baryons are 
 expressed  as
\begin{eqnarray}
& & <p \mid H^3 \mid p > = F+D \label{pfd}\\
& & <n \mid H^3 \mid n > =-( F+D) \label{nfd}\\
& & <\Sigma^+ \mid H^3 \mid \Sigma^+ > =
 F+D + \frac{1}{k}(F-D) \label{Sigmapfd}\\
& & <\Sigma^0 \mid H^3 \mid \Sigma^0 > = 0, \nonumber \\
& & <\Sigma^- \mid H^3 \mid \Sigma^- > =
 -(F+D) - \frac{1}{k}(F-D) \label{Sigmamfd}\\
& & < \Lambda \mid H^3 \mid \Lambda > = 0, \nonumber \\
& & <\Xi^0 \mid H^3 \mid \Xi^0 > \nonumber \\ 
& & = -\frac{1}{3}(F+D) - \frac{2}{3k}(F-D) - \frac{1}{k+2}
\{-3F+D - \frac{3}{k}(F-D)\} \label{Xi0fd}\\
& & <\Xi^- \mid H^3 \mid \Xi^- > \nonumber \\
& & = \frac{1}{3}(F+D) + \frac{2}{3k}(F-D) + \frac{1}{k+2}
\{-3F+D - \frac{3}{k}(F-D)\}. \label{Ximfd}
\end{eqnarray}
\par
As noted in the case of the matrix elements of generator $\lambda^8$
here we also find a systematic structure in the matrix elements of
generator $\lambda^3$. 
{}From these matrix elements $<B_f \mid H^3 \mid B_i >$ it is seen that
  the Coleman-Glashow relation 
is derived only for $N_c =3$. 
\par
Comparing these matrix elements of $H^3$ to those of $H^8$
we find special sets of the $F$ and $D$ terms which correspond to 
the $F/D$ ratios that  appear in the order $O(1)$, $O(1/N_c)$ and
$O(1/N_c^2)$  of $1/N_c$ expansion. 
The general structure of
the matrix elements of the flavor generators will be discussed elsewhere. 
 \par


\section{Summary}
In the preceding sections we have calculated the $F/D$ ratios of spin-nonflip
and spin-flip
 baryon vertices for arbitrary number of color degrees of freedom $N_c$
 both in the non-relativistic quark model
  with the  $SU(6)$
 spin-flavor symmetry and in  the chiral
 soliton model with  $SU(3)$ flavor symmetry. 
 We find that the spin-nonflip $F/D$ ratio of baryon vertex($F_+/D_+$)
  goes to $-1$ in the limit of
 $N_c \to \infty$, while that of  spinflip vertex($F_-/D_-$) goes
   to $1/3$.
 We have shown that the leading value  $F/D= -1$ of spin-nonflip baryon vertex
  in the $1/N_c$ expansion corresponds to the isoscalar dominance
  while the leading value $F/D=1/3$ of the spin-flip vertex obtained
  in our previous paper corresponds to the isovector  dominance,
  representing the $I=J$ rule for baryon vertices.
  \par
 We have derived the model independent result by using the NRQM and the CSM.
 These results can be interpreted as the $I=J$ rule for baryon vertex.
 \par
 We have explained the reason why the $I=J$ rule is satisfied.
 In the $SU(2)$ CSM, both of
 the isoscalar
 part of  spin-flip vertex  and  the isovector part of  
 spin-nonflip vertex contain  the  time-derivative of
 dymamical variable for the
 spin-isospin rotation ${\dot A}$, while the isovector part of
  spin-flip and the isoscalar part
 of spin-nonflip do not contain ${\dot A}$. 
 This is the content of the $I=J$ rule.
 \par
In terms of the matrix elements which transform as the generators $\lambda^8$  
and $\lambda^3$ of the $SU(3)$ symmetry we  derive the model independent
 isoscalar and isovector formula for baryon vertices and  apply these
 to the mass formulae and the isoscalar
 and isovector parts of the baryon magnetic moments.
 We obtain the same Okubo-Gell-Mann mass relation (\ref{og}) and the refined
  relation (\ref{rog}) 
among the octet baryons   
 as the one for the case $N_c=3$, derived model independently for
 arbitrary  color degrees of freedom $N_c$.
 In the case of   $\lambda^3$ of $SU(3)$
 symmetry the Coleman-Glashow mass relation is derived only for $N_c =3$.
\par
The matrix elements of generators $\lambda^8$ and $\lambda^3$
have a systematic structure. The leading part  in the $1/N_c$ expansion
gives common contributions to all states of baryon.
The second part, which is of the order $O(1/N_c)$ gives contributions
which increase as we go towards the  bottom of the root diagram of baryon
 states.
The third part appears only for states lying on the inner triangle
of the root diagram. We will discuss the general structure of
the matrix elements of the flavor generators elsewhere. 
 \par

\eject

\begin{flushleft}
Figure Captions 
\end{flushleft}
\vskip 5mm
Fig.~1. The root diagram for the $(k+1)(k+3)$ dimensional representation
 of baryon states. The ``octet" baryons are located in the upper part of
the root diagram.
\vskip 5mm
Fig.~2. $N_c$-dependence of  $F_-/D_-$ ratios of spin-flip baryon vertex
in the NRQM and CSM. The large $N_c$ limiting value  1/3 and the experimental 
value 0.58 $\pm 0.04$ which lies between the curves of NRQM and CSM.
\vskip 5mm
\noindent
Fig.~3. $N_c$-dependence of $D_+/F_+$ ratios of spin-nonflip baryon
vertex in the NRQM and CSM. Here $D/F$ ratios are shown instead of $F/D$
in order to avoid singular behaviors of the $F/D$ which appear 
at $N_c=3$ and $-2+\sqrt{13}$ in the NRQM and CSM, respectively. 
 The large $N_c$ limiting value $-1$ and the experimental 
value $-0.33 \pm 0.02$ which lies also
 between the curves  of NRQM and CSM are shown.

\newpage
\begin{center}
\epsfxsize=13.5cm
\hfil\epsfbox{Fig1.eps}\hfill
\end{center}

\newpage
\begin{center}
\epsfxsize=10cm
\hfil\epsfbox{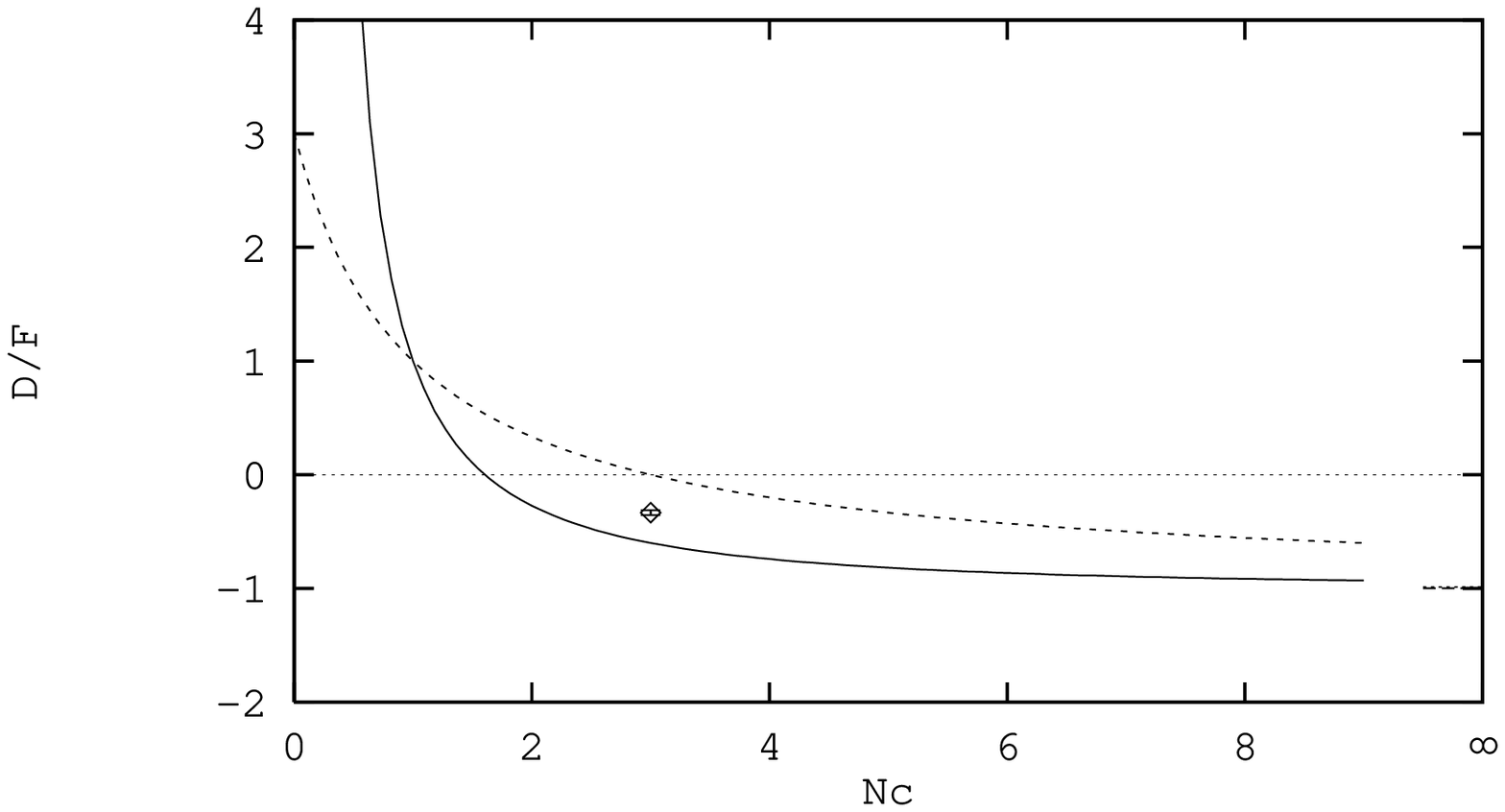}\hfill
\end{center}

\begin{center}
Fig.2 \vspace{1cm}\\
\end{center}

\begin{center}
\epsfxsize=10cm
\hfil\epsfbox{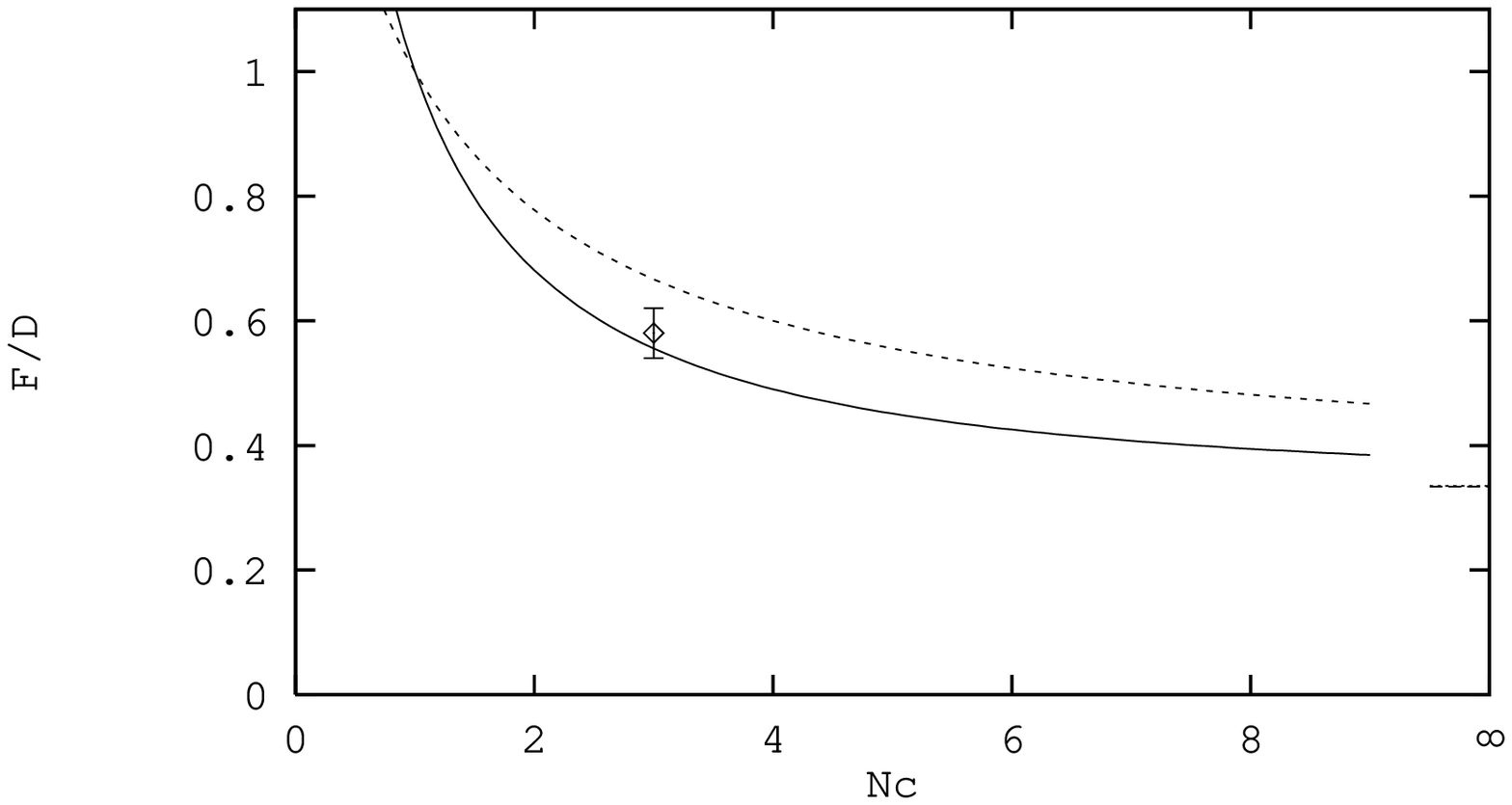}\hfill
\end{center}

\begin{center}
Fig.3
\end{center}

\end{document}